\documentclass[aps,twocolumn,letter,showpacs,preprintnumbers,amsmath,amssymb,superscriptaddress,bibnotes]{revtex4-1}

\usepackage{graphicx,epsfig}
\usepackage{dcolumn}
\usepackage{bm}
\usepackage{textcomp}
\usepackage{amsmath}
\usepackage{amssymb}
\usepackage{color}
\usepackage{soul}
\usepackage{natbib}

\begin{document}

\title{Controlling magnetoresistance by tuning semimetallicity through dimensional confinement and heteroepitaxy}

\author{Shouvik Chatterjee}
\email[Authors to whom correspondence should be addressed: ]{schatterjee@ucsb.edu, cjpalm@ucsb.edu, janotti@udel.edu}
\affiliation{Department of Electrical $\&$ Computer Engineering, University of California, Santa Barbara, CA 93106, USA}
\affiliation{Department of Condensed Matter Physics and Materials Science, Tata Institute of Fundamental Research, Homi Bhabha Road, Mumbai, 400005, India}

\author{Shoaib Khalid}
\affiliation{Department of Physics and Astronomy, University of Delaware, Newark, DE 19716, USA}
\affiliation{Department of Materials Science and Engineering, University of Delaware, Newark, DE 19716, USA}
\author{Hadass S. Inbar}
\affiliation{Materials Department, University of California, Santa Barbara, CA 93106, USA}\author{Taozhi Guo}
\affiliation{Department of Physics, University of California, Santa Barbara, CA 93106, USA}
\author{Yu-Hao Chang}
\affiliation{Materials Department, University of California, Santa Barbara, CA 93106, USA}
\author{Elliot Young}
\affiliation{Materials Department, University of California, Santa Barbara, CA 93106, USA}

\author{Alexei V. Fedorov}
\affiliation{Advanced Light Source, Lawrence Berkeley National Laboratory, Berkeley, California 94720, USA}

\author{Dan Read}
\affiliation{Department of Electrical $\&$ Computer Engineering, University of California, Santa Barbara, CA 93106, USA}
\affiliation{School of Physics and Astronomy, Cardiff University, Cardiff CF24 3AA, UK}

\author{Anderson Janotti}
\email[Authors to whom correspondence should be addressed: ]{schatterjee@ucsb.edu, cjpalm@ucsb.edu, janotti@udel.edu}
\affiliation{Department of Materials Science and Engineering, University of Delaware, Newark, DE 19716, USA}

\author{Christopher J. Palmstr\o m}
\email[Authors to whom correspondence should be addressed: ]{schatterjee@ucsb.edu, cjpalm@ucsb.edu, janotti@udel.edu}
\affiliation{Department of Electrical $\&$ Computer Engineering, University of California, Santa Barbara, CA 93106, USA}
\affiliation{Materials Department, University of California, Santa Barbara, CA 93106, USA}
\affiliation{California NanoSystems Institute, University of California, Santa Barbara, California 93106, USA}




\begin{abstract}

Controlling electronic properties via bandstructure engineering is at the heart of modern semiconductor devices. Here, we extend this concept to semimetals where, utilizing LuSb as a model system, we show that quantum confinement lifts carrier compensation and differentially affects the mobility of the electron and hole-like carriers resulting in a strong modification in its large, non-saturating magnetoresistance behavior. Bonding mismatch at the heteroepitaxial interface of a semimetal (LuSb) and a semiconductor (GaSb) leads to the emergence of a novel, two-dimensional, interfacial hole gas and is accompanied by a charge transfer across the interface that provides another avenue to modify the electronic structure and magnetotransport properties in the ultra-thin limit. Our work lays out a general strategy of utilizing confined thin film geometries and heteroepitaxial interfaces to engineer electronic structure in semimetallic systems, which allows control over their magnetoresistance behavior and simultaneously, provides insights into its origin.

\end{abstract}
\maketitle
\section{Introduction}

Semimetallic compounds offer an exciting platform to realize exotic quantum states of matter and novel material properties\cite{armitage2018weyl,ruan2016symmetry,chadov2010tunable,Li2015GdN}. Large, non-saturating magnetoresistance is one such example\cite{tafti2016resistivity,tafti2016temperature,liang2015ultrahigh,leahy2018nonsaturating,xu2017origin,zeng2016compensated,hu2008classical}, where spin-orbit coupling \cite{tafti2016resistivity,tafti2016temperature}, linearly dispersive states\cite{leahy2018nonsaturating}, charge compensation\cite{xu2017origin,zeng2016compensated}, and disorder effects\cite{hu2008classical} have been proposed as possible mechanisms for its origin. Though electronic structure is expected to play a key role, demonstration of controlling the magnetoresistance via electronic structure modification remains elusive, which might allow us to distinguish between the different proposed scenarios. To address this outstanding issue we fabricated epitaxial thin films of a semimetallic compound LuSb on GaSb substrates with different film thicknesses. Though LuSb is found to be a compensated semimetal in the bulk\cite{chatterjee2019weak,pavlosiuk2017fermi}, we establish that dimensional confinement differentially alters the occupation of electron and hole-like bands lifting carrier compensation. Loss of carrier compensation along with an overall reduction in carrier mobility in thinner films dramatically modifies their magnetoresistance behavior establishing the importance of carrier compensation. However, no evidence is found for the predicted semimetallic to semiconducting phase transition\cite{sandomirskii1967quantum} in dimensionally confined thin films of LuSb, which remain semimetallic in the ultra-thin limit.

Heteroepitaxial interfaces offer another potential avenue to control electronic properties in few atomic layer geometries and can lead to emergent ground states not realizable in the bulk\cite{stormer1999fractional,ohtomo2004high,sharan2019formation}. We show that the nature of the local coordination and chemical bonding at a heterointerface provides a novel route to realize a two-dimensional (2D) electron / hole gas. In particular, the heterointerface between a rock-salt (LuSb) and zinc-blende (GaSb) crystal structures results in the emergence of a 2D hole gas that remains tightly bound to the interface and is accompanied by a charge transfer across it, significantly affecting the electronic structure and transport properties in LuSb in the ultra-thin limit.


\section{Results and Discussion}


\subsection{Growth and Transport}

Epitaxial thin films of rock-salt LuSb (6.055\AA)\cite{przybylska1963preparation} were synthesized on a nearly lattice matched GaSb substrates (6.096\AA) with a thin GaSb epitaxial buffer layer using molecular-beam epitaxy (MBE). The c(2$\times$6) reconstruction of the Sb-rich GaSb(001) surface immediately changes to a (1$\times$1) reconstruction at the beginning of the deposition of LuSb atomic layers, shown in Fig.~1B. The LuSb atomic layers remain epitaxial and single phase with [001] out-of-plane orientation even in the ultra-thin limit, as revealed by the $\theta$-2$\theta$ x-ray diffraction scans shown in Fig.~1C. Details about further structural characterization of these films can be found in the Supplementary Materials (Section 1, Fig.~S1)\cite{suppl}. 

Dimensional confinement is found to significantly modify electrical properties of LuSb, which is a semimetal in the bulk with hole pockets ($\beta$ and $\delta$) at the zone center and an elliptical electron pocket at the zone edge ($\alpha$), as shown in Fig.~1A. While the 40, 32, and 20 ML thick films exhibit metallic behavior, a low temperature resistivity upturn can be seen in the 12, 8, and 6 ML thick films (Fig.~1D). At high temperatures, parallel conduction from the underlying GaSb buffer layers and GaSb substrate strongly influences temperature dependence of film resistance in the LuSb / GaSb thin films. Charge carriers in GaSb freeze out at low temperatures resulting in the film resistance being dominated by the LuSb atomic layers, which is the region of interest in this study (see Fig.~S2 in the Supplementary Materials\cite{suppl}). The temperature at which resistance of LuSb atomic layer starts to dominate depends on the relative resistance of the LuSb atomic layer in comparison to that of GaSb, which depends on the LuSb film thickness. To compare temperature dependence of the longitudinal resistance in thin films with different LuSb layer thicknesses we show normalized resistance R$_{xx}$/R$_{\Xi}$ in Fig.~1D, where R$_{\Xi}$ is the resistance below which the film resistance is dominated by the LuSb atomic layer as is evident from a dramatic change in slope as a function of temperature. R$_{\Xi}$ for different film thicknesses are shown by corresponding arrows in Fig.~1D. Magnetoresistance in these films also undergo dramatic changes both in magnitude and shape with film thickness (Fig.~1E). While the 40 ML thick film exhibits a MR of 120$\%$ at 14T, it drops down to less than 10$\%$ in the thinner films. Furthermore, the 8 ML and 6 ML thick films show a saturating behavior at high field values unlike the thicker films due to strong electron-electron interaction (EEI) effects (Fig.~S2 in the Supplementary Materials\cite{suppl}). Shubnikov de Haas oscillations (SdH) were observed in the magnetoresistance data, shown in Fig.~1E, for all the electronic bands ($\alpha$, $\beta$, $\delta$) for the 40 ML and 32 ML thick films, but only for the $\alpha$ and $\beta$ bands for the 20 ML thick film, and were absent in the thinner films indicating an overall reduction in mobility with decreasing film thickness (Figs.~1E,G). A smooth 5$^{th}$ order polynomial background was used to extract SdH oscillations from the magnetoresistance data, shown in Fig.~1G. Further details of the extraction procedure can be found in the Supplementary Materials (Section 3, Fig.~S3)\cite{suppl}. Indeed, quantum mobilities for the  $\alpha$ and $\beta$ bands were found to decrease by 84\% and 77\%, respectively, with the decrease in film thickness from 40 ML to 20 ML, shown in Fig.~1H. Details of the estimation of quantum mobilities is provided in the Supplementary Materials (Section 3, Fig.~S4)\cite{suppl}. The quantum oscillation measurements further indicate dimensional confinement induces a change in the electronic structure of LuSb, differentially affecting the electron and hole-like bands. The Fermi wavevector ($k_{F}$) for the hole-like $\beta$ pocket was found to have reduced from 0.1504 \AA$^{-1}$ to 0.1436 \AA$^{-1}$, a change of $\approx$ 0.007 \AA$^{-1}$, on reduction of the layer thickness from 40 ML to 20 ML. However, for the electron-like $\alpha$ pocket the change in the Fermi wavevector along the semi-minor axis was only from 0.112 \AA$^{-1}$ to 0.1107 \AA$^{-1}$, an insignificant change of $\approx$ 0.001 \AA$^{-1}$. This is also manifested in the Hall measurements where a multi-carrier Hall behavior in the 40 and 32 ML thick films changes to a single carrier electron-like behavior in the thinner films of 20 and 12 ML (Fig.~1F, also see Fig.~S3J in the Supplementary Materials), as evidenced from the linear Hall data for these films, up to a magnetic field of 14 Tesla. This can be understood as the predominance of electron-like carriers over hole-like carriers in thinner films, consistent with the results from SdH oscillation analysis, indicating that in thinner films electron-hole compensation is lifted with a much larger concentration of electron-like carriers. However, surprisingly in contrast to the thicker films, p-type Hall conductivity (positive slope of the Hall coefficient) is observed in the 8 and 6 ML thick films, which is discussed later. 


\subsection{ARPES}

To gain insights into our transport results we directly map out the evolution of the electronic structure of LuSb with decreasing film thickness by angle-resolved photoemission spectroscopy (ARPES). The measurements were taken at a photon energy of 60 eV, which samples a two-dimensional momentum region ($k_{x}$, $k_{y}$) of the three-dimensional Brillouin zone with $k_{z}$ close to the bulk $\Gamma$ point (see Fig.~2A) The Fermi surface of LuSb showing both the hole pockets ($\beta$, $\delta$) at the zone centre ($\Gamma$) and the elliptical electron pocket ($\alpha$) at the zone edge ($X$) is shown in Fig.~2B. In Fig.~2C-J, we show the hole and the electron pockets at the $\Gamma$ and the bulk X points ($\bar{\rm M}$ at the surface Brilluoin zone), respectively for different film thicknesses. The occupation of the hole pockets decreases dramatically as the film thickness is reduced, whereas no significant change in occupation of the electron pocket is observed for all but 6 ML thick film. For the 6 ML thick film we observe a slight reduction in ellipticity ($e = \frac{k_{F,semimajor}}{k_{F,semiminor}}$) of the electron pocket with $k_{F,semiminor}$ remaining unchanged from the bulk limit. Finite thickness of our films results in the formation of quantum well states observed for the hole-like carriers in the 20, 12, and 6 ML thick samples (Figs.~2C-F). We, however, do not observe the corresponding quantum well states for the electron pocket (Figs.~2G-J). This is most likely due to finite lateral coherence length in our thin films coupled with the fact that an off-normal geometry with a high polar angle had to be used in the ARPES measurements of the electron pocket due to their location at the zone-edge\cite{chiang2000photoemission} (see Fig.~1A,B). 

We estimated Fermi wavevectors ($k_{F}$) of all the Fermi surface sheets ($\alpha$, $\beta$, and $\delta$) from the ARPES data, shown in Fig.~2C-J, from where corresponding carrier concentration was estimated noting that the $\beta$ and $\delta$ bands are quasi-spherical ($n_{3D}$ = $\frac{k_{F}^{3}}{3\pi^{2}}$), while the $\alpha$ band is elliptical ($n_{3D}$ = $\frac{k_{F,semiminor}^{2}k_{F,semimajor}}{3\pi^{2}}$) (see Fig.~1A, 2B). Extracted $k_{F}$s for different film thicknesses are shown in Fig.~3B (also see Fig.~S5 in the Supplementary Materials). While the 40 ML thick film, which is in the bulk limit, is a compensated semi-metal with near equal concentration of electron ($n$ = $3.22\times10^{20}$ cm$^{-3}$) and hole-like carriers ($p$ = $3.18\times10^{20}$ cm$^{-3}$) with a ratio $\frac{n}{p}$ $\approx$ 1.01, thinner films are not carrier compensated with the effect being exacerbated as the film thickness is reduced. For the 6 ML thick film we observe two occupied sub-bands for the $\delta$ pocket and a single sub-band for the $\beta$ pocket with an effective hole-like sheet carrier concentration of 5.97$\times10^{13}$ cm$^{-2}$. Even considering finite occupation for only the lowest sub-band of just the in-plane electron pockets (i.e. those along $k_{x}$ and $k_{y}$, with confinement along $k_{z}$, see Fig.~1A) we obtain electron-like sheet carrier concentration of 9.58$\times10^{13}$ cm$^{-2}$ with the ratio $\frac{n}{p}$ $\approx$ 1.6, which is a conservative lower bound, but is still far away from electron-hole compensation. Hence, the modification in the magnetoresistance behavior in LuSb thin films can be directly ascribed to the loss of electron-hole compensation in thinner atomic layers. Furthermore, no band inversion is observed in our measurements showing that topological aspects of the band structure are not important for the magnetoresistance in LuSb. 

To understand the origins of the thickness dependent changes of the electronic structure observed in our thin films we performed slab calculations of LuSb using density functional theory (DFT). Details of the calculations can be found in the methods and in the Supplementary Materials (Section 4, Figs.~S7-S9)\cite{suppl}. Our calculations predict lifting of the electron-hole compensation in thinner films due to a differential reduction in the occupation of the electron and hole-like carriers with film thickness, in accordance with our experimental observation. However, the electron pocket ($\alpha$), at $\bar{\rm M}$ is also strongly affected by quantum confinement in our calculations (Fig.~3A), contrary to the experimental data shown in Fig.~3B. This apparent discrepancy stems from the additional interfacial effects and the resulting charge transfer into the LuSb atomic layers across the LuSb/GaSb interface (see Fig.~3C), presence of which is revealed in our slab calculations when the substrate, GaSb is also included, as discussed in the next section.


\subsection{Role of the interface}

While the ARPES and quantum oscillation measurements along with the LuSb slab calculations establish the loss of electron-hole compensation in thinner films, dimensional confinement alone is not sufficient to explain all aspects of our experimental data. First, our observation of a positive Hall coefficient in the thinner films of 8 and 6 ML is in apparent contradiction to the ARPES and quantum oscillation measurements that show a larger carrier concentration of the electron-like carriers compared to hole-like carriers in these films. Second, evolution of the electron-like $\alpha$ pocket with film thickness could not be reproduced by slab calculations when only LuSb atomic layers were considered, as noted earlier. 

To understand the origins of these seemingly contradictory behavior we evaluate the nature of the interface in our thin film heterostructure. A change in sign of the Hall coefficient from negative to positive values in thinner films could arise if the interface between LuSb and GaSb hosts a two-dimensional hole gas, which will become increasingly important, and manifest in transport measurements in the ultra-thin limit having a contribution comparable to that of the bulk film. To explore this scenario we have fitted the transport data with a three-component Drude conductivity model,
\begin{multline}
    G_{xx} = \frac{R_{xx}}{R^{2}_{xx}+R^{2}_{xy}} =  en_{e}\mu_{e}t + en_{h}\mu_{h}t + en_{2D}\mu_{2D}\\
    G_{xy} = \frac{R_{xy}}{R^{2}_{xx}+R^{2}_{xy}} =  B\times[-en_{e}\mu^{2}_{e}t + en_{h}\mu^{2}_{h}t + en_{2D}\mu^{2}_{2D}],
\end{multline}
where the first and the second terms on the right hand side represent contributions from the bulk of the film for the electron and hole-like carriers, respectively. The third term represents contribution from the interfacial state. $n$, $\mu$, and $t$ represent carrier concentration, mobility and film thickness, respectively. $G_{xx}$ and $G_{xy}/B$ follows approximately $t^{2}$ and $t^{3}$ behavior, respectively, between 12 and 32 ML and saturates for film thicknesses less than 12 ML, as shown in Fig.~4B,C. The value to which the conductivity value saturates represents the contribution of the interfacial charge carriers from which we obtain an interfacial carrier concentration of $n_{2D}$ = 4.38$\times$10$^{14}$ cm$^{-2}$ ($\approx$0.8 holes/2D unit cell) and a mobility of $\mu_{2D}$ = 1.05 cm$^{2}$/Vs. Furthermore, our analysis indicates that the mean free path, and hence the mobility of the charge carriers in the bulk of the film for film thicknesses between 12 and 32 ML is primarily dominated by interfacial scattering and is thus proportional to the film thickness $t$. The 40 ML thick film deviates from this trend suggesting that at this thickness the film is in the bulk 3-D limit, which is in accordance with the absence of quantum well states in the ARPES measurements. 

The existence of a two-dimensional hole gas at the LuSb/GaSb (001) interface is revealed in our DFT slab calculations when the substrate GaSb is included. In Fig.~5A we show the electronic band structure of the LuSb/GaSb/LuSb (001) slab. We find two interfacial bands crossing the Fermi level, mostly composed of $s$ orbital of the Ga atoms at the interface, corresponding to one band with Ga-$s$ character per interface.  These two bands show different dispersions along $\Gamma$-$\bar{\rm X}_1$-$\bar{\rm M}$ and $\Gamma$-$\bar{\rm X}_2$-$\bar{\rm M}$ because the corresponding Ga-Sb bonds at the two equivalent interfaces are rotated with respect to each other by 90$^{\circ}$ due to the symmetry of the zinc blende structure of GaSb. The Fermi level crossings of these two bands and the resulting Fermi surface is shown in Fig.~5A and 5D, respectively, indicating a hole-like behavior, which explains experimental observation. The estimated carrier concentration, based on the Luttinger volume is 0.7 hole per 2D unit cell area per interface. We note that the estimated Luttinger volume depends marginally on the functional used in the calculation, and should be considered as a lower bound due to the use of GGA in the slab calculations (see Methods). The calculated projected density of states, shown in the Supplementary Materials (Fig.~S9)\cite{suppl}, also indicates that these interface bands are associated with the Ga atoms at the interface. The charge density distribution (Fig.~5B) corresponding to the square of the single-particle states at the maximum of the hole band along $\bar{\rm X}_2$-$\bar{\rm M}$ (highlighted by blue circles in Fig.~5A), reveals the two-dimensional character of these two bands, which are highly localized out-of-plane, yet uniformly distributed in the plane of the interface. Our calculations also predict distortion of the Lu atoms at the interface due to bonding mismatch, shown in  Fig.~5E. This is directly observed in the high-angle annular dark field scanning transmission electron microscopy (HAADF-STEM) data (see Figs.~5E-G and Fig.~S10 in the Supplementary Materials \cite{suppl}), showing that the Lu-Lu inter-atomic distance in the out-of-plane direction is smaller at the interface, consistent with the predicted atomic layer buckling. This provides further evidence for the validity of our slab calculations in understanding the experimental results.

The charge density of this two-dimensional hole gas and associated charge transfer to the LuSb atomic layers can also be estimated based on a simple electron counting argument. In Fig.~5C, we show the valence charge density profile of the  LuSb/GaSb/LuSb (001) slab system.  The excess charge on the LuSb layer on the top or at the bottom of the GaSb layer is defined as the macroscopically averaged charge density along the heterostructure direction in the LuSb region (indicated in Fig.~5C) and the corresponding charge density of charge neutral LuSb bulk.  The calculated excess charge density on the LuSb film amounts to 0.45 electrons/2D interface unit cell area. Assuming that an excess charge of 1.5 electrons is expected at the interface due to the valence mismatch, i.e. in an ionic picture of GaSb, each Ga layer transfers 3/2 electrons to each of the neighbouring Sb layers, we obtain $1.5-0.45=1.05$ electrons/2D unit cell area would remain at the interface.  Discounting the electrons that are transferred to the bulk of GaSb, which is only 0.045 electrons/2D unit cell area per interface according to the results shown in Fig.~5C, this will amount to $\sim$1 excess electrons per 2D unit-cell area per interface. Assuming that this electron partially occupies an interface band, we have about 1 hole per unit-cell area in the interface band. This estimation of 1 hole/2D unit cell can be considered as an upper bound value as GaSb bonds have strong covalent character.  

The estimated range of $0.7-1$ hole per 2D unit-cell, based on the DFT calculations, corroborates the experimental value of 0.8 hole/2D unit cell for the interface conducting channel obtained from the analysis of the Hall conductivity above. Furthermore, following our theoretical understanding, observation of 0.8 hole/2D unit cell at the interface implies 0.25 excess electrons transferred to the LuSb atomic layers. Calculated $k_{F}$ values for the LuSb slabs after inclusion of the charge transfer across the interface were found to be in close agreement with those extracted from ARPES, shown in Fig.~3C. This underscores the importance of the interfacial effects, in addition to the quantum confinement effects, in understanding the evolution of the electronic structure with film thickness in heteroepitaxial semimetallic thin films. 

We note that an alternate explanation for the 2D hole gas at the LuSb/GaSb (001) interface could be due to band bending and surface accumulation in the GaSb interfacial layer resulting from the Fermi level at the GaSb surface being pinned in the valence band. We rule this out for the following reasons. First, the surface Fermi level pinning position of uncovered GaSb surfaces is known to lie 0.2 eV above the valence band\cite{kudrawiec2012contactless}. Second, both the photoemission results from metal/GaSb heterostructures and electrical measurements of metal/GaSb Schottky barriers indicate a surface Fermi level position within the GaSb band gap\cite{chye1978photoemission,chye1978evidence,nishi2014study,murawala1990barrier,rotelli1997photoelectric}. Finally, the density of such electrostatically induced 2D hole gases is typically $\leq$5$\times$10$^{12}$ holes/cm$^{2}$\cite{altarelli1987electronic,mokerov1999high}, which is two orders of magnitude lower than the hole density observed in our experiments. Therefore, the 2D hole gas observed in our experiments is of novel origin arising due to bonding mismatch at the interface, which should be a generic feature of such heterointerfaces.


\subsection{Semimetal to semiconductor transition}

 The ARPES measurements shown in Fig.~2 clearly reveal that a finite occupation is maintained for both the electron and hole like bands, thereby preserving the semimetallic character of LuSb even at the ultra-thin limit of 6 ML. This is at odds with the resistivity upturn at low temperatures observed in the transport measurements of the thinner films (Fig.~1D).  In Figs.~6A and 6B we show that the conductance drops logarithmically with temperature for the 6 and 12 ML thick films, respectively.
 \begin{equation}
 \Delta\/G(T) = \frac{e^{2}}{\pi h}Aln(\frac{T}{T_{0}})
 \end{equation}
 where $T_{0}$ is the reference temperature. This can arise from quantum interference (QI) effects such as weak localization and electron-electron interaction (EEI) in the two-dimensional limit \cite{lee1985disordered,altshuler1980interaction}. To distinguish between the two cases we examine the change in slope in the temperature dependence of conductance as a function of applied magnetic field. 
 QI effects can be readily suppressed on application of magnetic field, while EEI effects are more robust due to larger characteristic fields \cite{lee1985disordered,altshuler1985electron}. Therefore, the pre-factor A in eqn. 2 obtained at high field values $A_{high}$ is solely due to EEI effects $A_{high} = A_{ee}$, whereas the one at zero field is a combination of both QI ($A_{QI}$) and EEI effects ($A_{ee}$), $A_{low} = A_{ee} + A_{QI}$. We estimate $A_{QI}$ to be -0.07 and -0.15 and $A_{ee}$ equal to 0.44 and 0.39 for the 12 and 6 ML thick film, respectively. We obtain a negative value for A$_{QI}$ for both the 12 and 6 ML thick film, consistent with the observation of weak anti-localization (WAL) behavior in magnetoresistance, shown in Figs.~6D-F.  Angular dependence of magnetoresistance confirms the two-dimensional character of WAL, while further distinguishing it from the effects of Zeeman splitting on the EEI correction that also results in a dip in the zero-field magnetoresistance, but is insensitive to the magnetic field direction at low fields \cite{lee1985disordered,altshuler1985electron}. Having established that the low field magnetoresistance behavior is dominated by weak antilocalization effect, we investigate the scattering mechanisms in these films utilizing the Hikami-Larkin-Nagaoka (HLN) theory\cite{hikami1980spin} applicable for a two-dimensional electronic system, given by:
\begin{align}
 \Delta G_{WAL}(B) & = \alpha \frac{e^{2}}{\pi h}[\frac{1}{2}(\Psi(\frac{1}{2} + \frac{B_{\phi}}{B}) - ln(\frac{B_{\phi}}{B})) \nonumber \\
 & - \frac{3}{2}(\Psi(\frac{1}{2} + \frac{\frac{4}{3}B_{SO} + B_{\phi}}{B}) - ln(\frac{\frac{4}{3}B_{SO} + B_{\phi}}{B}))],
 \end{align}
where $B_{\phi}$ and $B_{SO}$ are the characteristic fields corresponding to inelastic and spin-orbit scattering, respectively, and $\Psi(x)$ is the digamma function. Fits to the magnetoconductance data is shown in Fig.~6E and~6F, and extracted temperature dependence of the characteristic dephasing field $B_{\phi}$ for the 12 and 6 ML films are shown in Figs.~6G and~6H, respectively. At 2 K, the phase coherence lengths ($l_{\phi}$) were found to be 193 nm and 47 nm and the spin-orbit scattering lengths ($l_{SO}$) 11 nm and 10 nm for the 12 and 6 ML thick films, respectively\cite{suppl}. Temperature dependences of $l_{\phi}$ and $l_{SO}$ are shown in Fig.~S6A in the Supplementary Materials. Temperature dependence of $B_{\phi}$ can be well approximated as 
\begin{align}
B_{\phi} & = B_{0} + B_{ee} + B_{eph} = a + bT + cT^{n},
\end{align}
where $B_{0}$, $B_{ee}$, and $B_{eph}$ are contributions due to impurity, electron-electron, and electron-phonon scattering, respectively, $a$, $b$, and $c$ are constants, and $n$ varies between 2 and 4 \cite{lin2002recent}. For the 12 ML thick film at low temperatures, electron-electron scattering dominates, which is quickly overshadowed by contributions from the electron-phonon scattering at higher temperatures exhibiting a T$^{2.2}$ dependence. However, for the 6 ML thick film, $B_{\phi}$ shows a linear temperature dependence suggesting the predominance of EEI effects in thinner films and also much weaker electron-phonon scattering. Our observation thus explains the origin of resistivity upturn at low temperatures in these films despite the absence of a bulk band gap.


\section{Conclusion}

In summary, we have shown how magnetoresistance behavior can be modified and charge compensation can be lifted in an otherwise compensated semimetallic system by dimensionally confining charge carriers in ultra-thin films. This approach has allowed us to distinguish the underlying mechanism behind the observed magnetoresistance behavior while establishing the efficacy of few atomic layer geometries in controlling electronic properties, which can be readily extended to other semimetallic systems. Though the presence of surface states at the interface between a rock salt and a zinc blende crystal structure\cite{bomberger2016growth,krivoy2012growth,driscoll2001electronic} had been speculated  in earlier studies, our experimental and theoretical work established the presence of a two-dimensional hole gas at this technologically relevant heterointerface\cite{palmstro1990lattice} and also elucidated its origin, which significantly affects the electronic and transport properties in the ultra-thin limit. We have shown that controlling the nature of chemical bonding at the interface offers a novel route to realize 2D hole gas, which is distinct from a traditional 2D hole gas that arises due to the formation of accumulation layer near the interface as a result of band bending in semiconductors. Such emergent interfacial 2D hole/electron gas is expected to be generic in heterointerfaces with different bonding configurations and can profoundly influence advanced device geometries including those currently under investigation for topological quantum computing\cite{liu2019semiconductor}. No evidence for the lifting of semimetallicity is found in these thin heteroepitaxial films. However, our analysis of quantum interference and electron-electron interaction effects establish the inadequacy of transport measurements alone in understanding either the predicted semimetallic to semiconducting phase transition or the evolution of the electronic structure with film thickness. Our work provides a comprehensive understanding of the electronic structure in ultra-thin semimetallic systems that will be important in possible device applications\cite{bomberger2017overview} and in the realization of novel physical properties that are proposed to emerge in the ultra-thin limit\cite{ruan2016symmetry,Li2015GdN,khalid2019topological,inoue2019band}. Our work also sets the stage for further control over their electronic properties by applications of bi-axial stress and proximity effect in artificial heterostructures.


\section*{Acknowledgments}

Synthesis of thin films, development of the UHV suitcase, ARPES experiments, and theoretical work were supported by the US Department of Energy (Contract no.~DE-SC0014388). Development of the growth facilities and low temperature magnetotransport measurements were supported by the Office of Naval Research through the Vannevar Bush Faculty Fellowship under the Award No. N00014-15-1-2845. Scanning probe studies were supported by National Science Foundation (Award number DMR-1507875). This research used resources of the Advanced Light Source, which is a DOE Office of Science User Facility under contract no.~DE-AC02-05CH11231. We acknowledge the use of shared facilities of the National Science Foundation (NSF) Materials Research Science and Engineering Center (MRSEC) at the University of California Santa Barbara (DMR 1720256) and the LeRoy Eyring Center for Solid State Science at Arizona State University. A portion of this work was performed in the UCSB Nanofabrication Facility, an open access laboratory. Density functional theory calculations made use of the National Energy Research Scientific Computing Center (NERSC), a U.S. Department of Energy Office of Science User Facility operated under Contract no.~DE-AC02-05CH11231. D.R. gratefully acknowledges support from the Leverhulme Trust via an International Academic Fellowship (IAF-2018-039).


\section*{Author Contributions}
S.C. and C.J.P. conceived the study. Thin film growth, film characterization, and transport measurements were performed by S.C. with assistance from H.S.I., T.G., and D.R. DFT calculations were performed by S.K. under the supervision of A.J. ARPES measurements were performed by S.C. with assistance from H.S.I, D.R, Y-H.C., E.Y., and A.V.F. and analyzed by S.C. TEM measurements and analysis were performed by A.G. The manuscript was prepared by S.C., S.K., A.J., and C.J.P. All authors discussed results and commented on the manuscript.
\\
\section*{Data Availability}
All data needed to evaluate the conclusions in the paper are present in the paper and/or the Supplementary Materials. Additional data supporting the findings of this study can be obtained from the corresponding author(s) upon request.
\\
\section*{Competing Interests}
The authors declare no competing interests.

\section*{Materials and Methods}

\subsection{Growth}

Thin films were grown by molecular-beam epitaxy (MBE) in a MOD Gen II growth chamber. A 5 nm-thick GaSb buffer layer was grown on low n-type doped GaSb (001) substrates (carriers freeze out at low temperatures, see Fig.~S2A in the Supplementary Materials) at 450$^{\circ}$C under Sb$_{4}$ overpressure after desorption of the native oxide using atomic hydrogen. This is followed by co-evaporation of Lu and Sb from calibrated effusion cells with the substrate temperature at 380$^{\circ}$C and the Lu to Sb flux ratio 1:1.10. The atomic fluxes of Lu and Sb are calibrated by Rutherford backscattering spectrometry (RBS) measurements of the elemental areal density of calibration samples on Si. These measurements were used to calibrate \emph{in-situ} beam flux measurements using an ion gauge. The sample surfaces were protected with a 5-nm-thick AlOx layer using e-beam evaporation before taking them out of the UHV chamber. For ARPES measurements conductive n-type Te doped GaSb (001) substrates were used. Similar growth procedure was followed in our earlier work, as described in \cite{chatterjee2019weak}

\subsection{HAADF-STEM}

For the structural analysis using scanning transmission electron microscopy (STEM), cross-sectional lamellas were prepared using a FEI Helios Dual-beam Nanolab 650 focused gallium ion beam (FIB) system. A 3 $\mu$m thick platinum(Pt) layer was deposited on the surface of the sample as a protective layer. Thereafter, FIB milling steps down to 2 kV were used to polish the lamella to approximately 50 nm in thickness. To minimise oxidation of the sample, the lamella was then immediately transferred to the STEM system. The high-angle annular dark field STEM (HAADF-STEM) imaging was carried out in a ThermoFisher Talos G2 200X TEM/STEM system using the ThermoFisher Scientific Velox software. The lamella was imaged along the GaSb [110] zone axis. A series acquisition was performed with a dwell time of 200 ns and Drift Corrected Frame Integration (DCFI) was used to process the acquired images. Quantification of distances between atomic peaks in the STEM image was done using ImageJ and MATLAB software. Line intensity profiles (widths integrated over 5 pixels) were acquired at the center of the required atomic columns. Intensity profiles over 10 -15 columns, having the same atomic configuration, were averaged to get final atomic profiles with high signal to noise ratio. Gaussian peak fitting was used to identify peak positions and calculate Lu-Lu or Sb-Sb atomic distances.

\subsection{Transport Measurements}

Following similar procedure described in our previous work \cite{chatterjee2019weak}, transport measurements were performed on fabricated Hall bar devices using standard a.c. lock-in technique  at low temperatures with the current flowing along [110] crystallographic direction, where parallel conduction from the substrate and the buffer layers can be neglected at low temperatures (see Fig.~S2A-C in the Supplementary Materials). The Hall bars were fabricated using standard optical lithography, followed by an ion milling procedure using argon ions. The contacts were made using 50 $\mu$m gold wire bonded onto gold pads. Low temperature measurements were carried out in a Quantum Design PPMS with base temperature of 2 K and maximum magnetic field of 14 T.

\subsection{ARPES}

Samples were transferred in a custom-built vacuum suitcase from the growth chamber at Santa Barbara to the ARPES endstation 10.0.1.2 at the Advanced Light Source in Berkeley. The pressure inside the vacuum suitcase was better than 1$\times$10$^{-10}$ Torr. Tunable synchrotron light in the 20 - 80 eV range was used for the photoemission measurements with a Scienta R4000 analyzer. The base pressure of the analysis chamber was better than 5$\times$10$^{-11}$ Torr. Similar sample transfer and ARPES measurement conditions were followed in our earlier work, as described in \cite{chatterjee2019weak}

\subsection{DFT Calculations}

First-principles calculations, based on the density functional theory (DFT) and projector augmented wave (PAW)\cite{blochl1994projector} method as implemented in the VASP code\cite{kresse1993ab,kresse1994ab}, were carried out to study the electronic structure of the LuSb/GaSb (001) interface. For the exchange and correlation we employed the generalized gradient approximation (GGA) of Perdew-Burke-Ernzerhof\cite{perdew1996generalized}. Test calculations based on the screened hybrid functional HSE06\cite{heyd2003hybrid,HSE} were used to overcome the problem of DFT-GGA in overestimating the overlap of the electron and hole pockets in LuSb\cite{khalid2020hybrid, chatterjee2019weak} and underestimating the band gap of GaSb.
The effects of spin-orbit coupling (SOC) are included in all band structure calculations. More details on the calculations can be found in the Supplementary Materials (Section 4, Figs.~S7-S9)\cite{suppl}.
The LuSb/GaSb (001) interface was simulated using (i) a slab geometry with LuSb/GaSb/LuSb layers with 7.5 layers of Ga-terminated GaSb sandwiched between two 6-monolayer thick LuSb with a $\sim$15 \AA\/ thick vacuum layer (Fig.~5), and (ii) a LuSb/GaSb (001) superlattice with 17 layers of LuSb and 7.5 layers of GaSb without any vacuum layer (Fig.~S9 in the Supplementary Materials). In both the cases there are two equivalent LuSb/Ga-terminated interfaces, which are rotated by 90$^\circ$ with respect to each other. 
LuSb (001) thin films were simulated using periodic slabs with 7, 13, 21, and 41 monolayers (ML). The odd numbers of layers are chosen to ensure inversion symmetry, making it easier to analyze the bandstructures. These calculations were performed using the DFT-GGA functional with 12$\times$12$\times$1 special $k$-points; HSE06 calculations for these slabs are prohibitively expensive given the size of the supercell and the large number of $k$-points required to describe metallic systems. The results are shown in Fig.~S8 in the Supplementary Materials. We note that the electron pocket at $\Gamma$ seen in the bulk (Fig.~S7) and in the surface band structures (Fig.~S8) are projections of the electron pockets at the $X_3$ in the bulk primitive cell (2 atoms/per cell) that is folded to $\Gamma$ point when using the 4-atom tetragonal unit cell of LuSb (001) and the slabs.

\section*{Supplementary Materials}

Supplementary Materials contain four sections and ten figures.



\begin{figure*}
\includegraphics[width=1\textwidth]{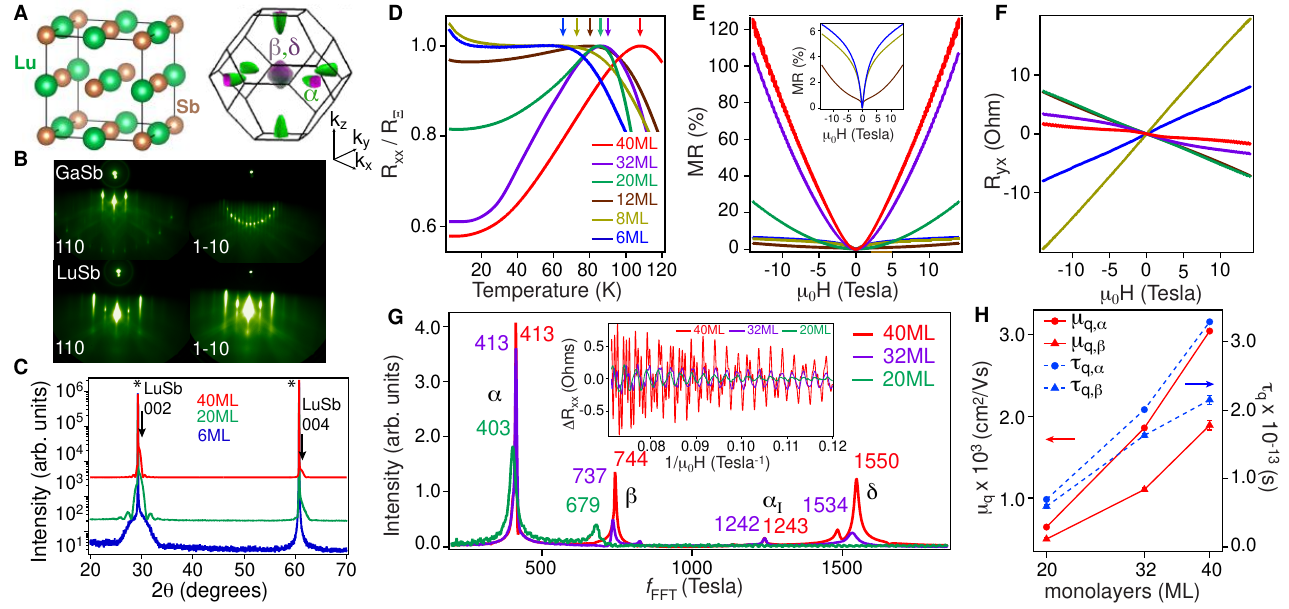}
\caption{\textbf{Synthesis and transport properties of LuSb/GaSb (001) thin films.} (A) Crystal structure of LuSb and its Fermi surface, calculated using hybrid DFT (B) RHEED images along the [110] and [1-10] azimuths. (C) Out-of-plane $\theta$-2$\theta$ XRD scans for epitaxial films of different thicknesses studied in this work. Substrate peaks are marked by asterisks. XRD scans are offset in intensity (along the y-axis) for clarity. (D) Temperature dependence of resistance in thin films of various thicknesses. R$_{\Xi}$ is the resistance at the sample temperature below which film resistance is dominated by LuSb layer. Temperatures corresponding to R$_{\Xi}$ are indicated for all film thicknesses. (E) Evolution of magnetoresistance with film thickness. Inset highlights saturating magnetoresistance behavior at high fields for 8 ML and 6 ML thick samples. All data taken at 2K. (F) Hall resistance measured at 2K as a function of film thickness. (G) Fast Fourier Transform (FFT) of the quantum oscillations for the 40, 32, and 20 ML thick films. Corresponding resistance oscillations are shown in the inset. $\beta$ and $\delta$ are the hole pockets at the zone center. $\alpha$ and  $\alpha_{I}$ are the frequencies corresponding to the projection of the elliptical electron pockets along the magnetic field direction ($k_{z}$) and those lying in the plane perpendicular to it ($k_{x}$, $k_{y}$), respectively, as shown in \textbf{A}. (H) Extracted quantum mobility and carrier lifetime for the $\alpha$ and $\beta$ pockets for the 40, 32, and 20 ML thick films.}
\label{fig:Transport}
\end{figure*}

\begin{figure*}
\includegraphics[width=1\textwidth]{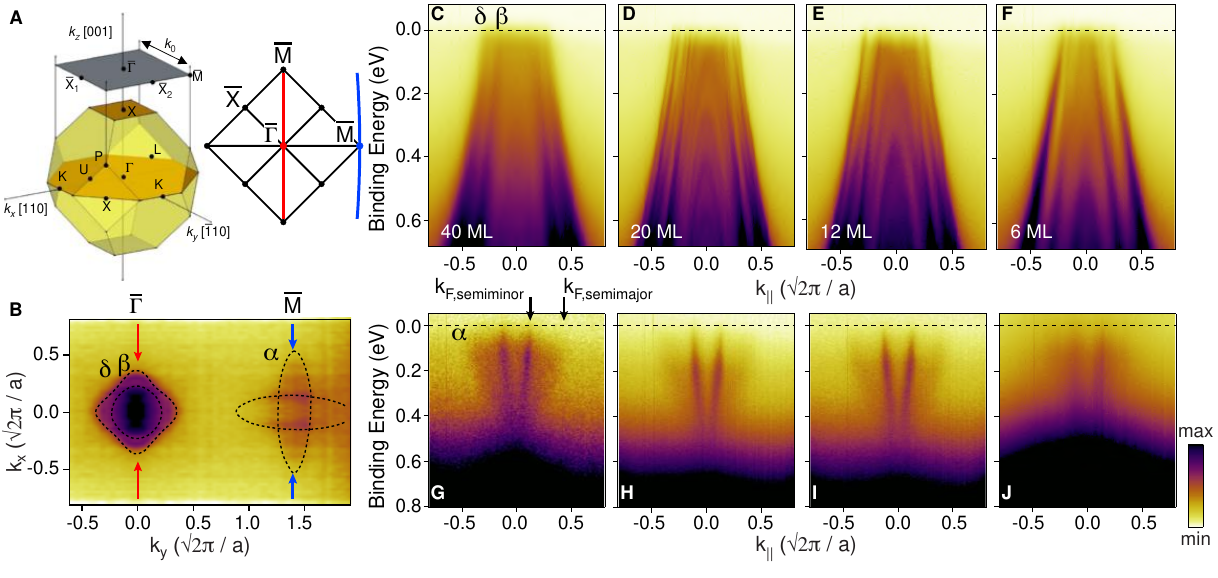}
\caption{\textbf{Photoemission spectroscopy of LuSb/GaSb (001) thin films.} (A) Three-dimensional and surface Brillouin zone of LuSb showing the high-symmetry points. Red and blue lines show the cut directions along which ARPES measurements are taken for \textbf{C-F} and \textbf{G-J}, respectively. (B) Fermi surface map of bulk LuSb\cite{chatterjee2019weak} showing both the electron and the hole pockets and the ARPES cut directions. Calculated Fermi surface obtained from DFT using screened hybrid functional (HSE06) is shown by black dotted lines. $E$-$k$ spectral map at the bulk $\Gamma$ point (top panels) along $\bar{M}$ - $\bar{\Gamma}$ - $\bar{M}$(red line in \textbf{A}) of the surface Brillouin zone for thin films of thickness (C) 40 ML (D) 20 ML (E) 12 ML and (F) 6 ML, and at the bulk $X$ point (bottom panels) along $\bar{\Gamma}$ - $\bar{M}$ - $\bar{\Gamma}$(blue line in \textbf{A}) of the surface Brillouin zone for (G) 40 ML (H) 20 ML (I) 12 ML and (J) 6 ML thick films. All data taken at 70 K and a photon energy of 60 eV.}
\label{fig:ARPES}
\end{figure*}

\begin{figure}
\includegraphics[width=1\columnwidth]{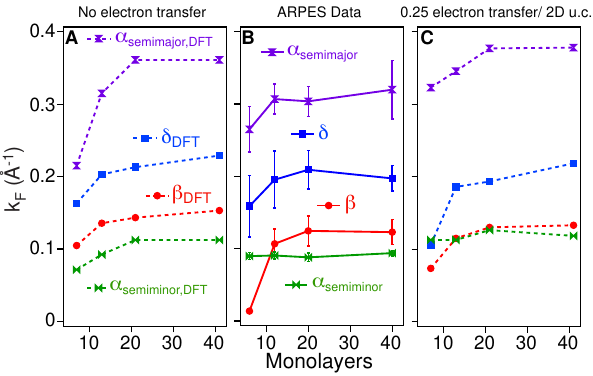}
\caption{ \textbf{Evolution of Fermi wavevectors ($k_{F}$) with film thickness.} Fermi wavevectors for both the elliptical electron-like pocket ($\alpha$) and quasi-spherical hole-like pockets ($\beta,\delta$) as a function of film thickness obtained from (A) LuSb slab calculation without considering charge transfer across the LuSb/GaSb interface, (B) ARPES measurements and (C) LuSb slab calculations after adding a charge transfer of 0.25 electrons/2D surface unit cell into the LuSb atomic layers (see text).}
\label{fig:kF}
\end{figure}

\begin{figure}
\includegraphics[width=1\columnwidth]{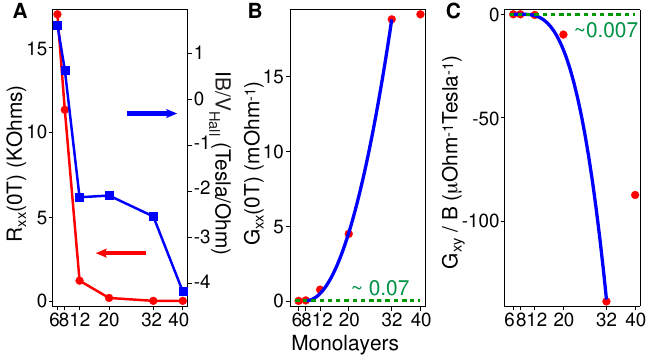}
\caption{ \textbf{Two-dimensional hole gas at the LuSb/GaSb interface.} (A) Longitudinal resistance ($R_{xx}$) at zero magnetic field and inverse of the slope in the Hall data, shown in Fig.~\textbf{1F}, as a function of film thickness. (B) Longitudinal conductance ($G_{xx}$) and  (C) Transverse conductance coefficient ($G_{xy}/B$), as defined in the text, as a function of film thickness. Saturation values of $G_{xx}$ and $G_{xy}/B$ are shown by green dotted lines.}
\label{fig:Interface}
\end{figure}

\begin{figure*}
\includegraphics[width=1\textwidth]{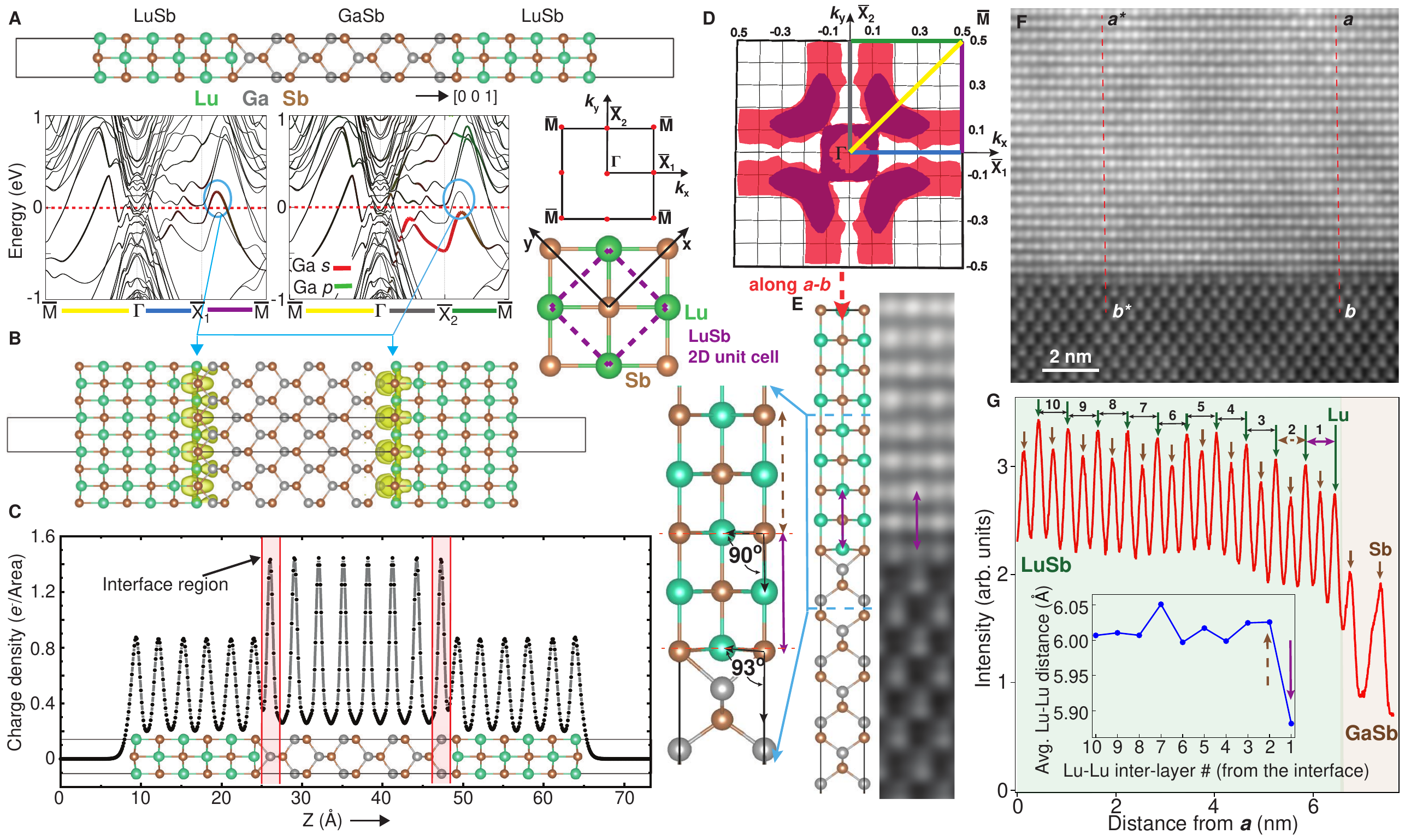}
\caption{ \textbf{Origin of the two-dimensional hole gas at a rock salt/zinc blende interface.} Electronic structure of a LuSb/GaSb/LuSb(001) slab showing (A) Interface bands that originate from  Ga $s$-orbitals at the interface. The two-dimensional unit cell and the surface Brillouin zone is also shown. (B) Charge density distribution of the single-particle state at the maximum of the hole bands crossing the Fermi level along the $\bar{\rm X}_1$-$\bar{\rm M}$ and $\bar{\rm X}_2$-$\bar{\rm M}$ directions. (C) Valence charge density distribution of the LuSb/GaSb/LuSb (001) slab along the direction perpendicular to the interface. (D) Fermi surfaces of the two interface bands associated with the two equivalent interfaces, yet rotated, corresponding to the structure used in the simulation, shown in \textbf{A}. (E) A closeup of the HAADF-STEM image in \textbf{F} and a ball-and-stick model of the proposed interface showing the atomic arrangement at the interface. (F) HAADF-STEM image along [110] zone-axis.  (G) Average atomic positions along \textit{\textbf{a}}-\textit{\textbf{b}}, shown in \textbf{F}, plotted as a function of the distance from \textit{\textbf{a}}, calculated by averaging those atomic columns that terminate with a Lu atom at the interface between \textit{\textbf{a}}-\textit{\textbf{b}} and \textit{\textbf{a*}}-\textit{\textbf{b*}}. Inset shows the evolution of the Lu-Lu inter-atomic distance as a function of the distance from the interface. Lu-Lu inter-atomic pairs are numbered away from the interface, as shown. Distortion of the Lu atoms at the interface is evident from the change in bond angle in the calculations, shown in \textbf{E}, which results in a reduced Lu-Lu inter-atomic distance between the 1st and the 3rd LuSb monolayer (violet arrow) compared to the 3rd and the 5th monolayer (brown dashed arrow).}
\label{fig:DFT-interface}
\end{figure*}

\begin{figure*}
\includegraphics[width=1\textwidth]{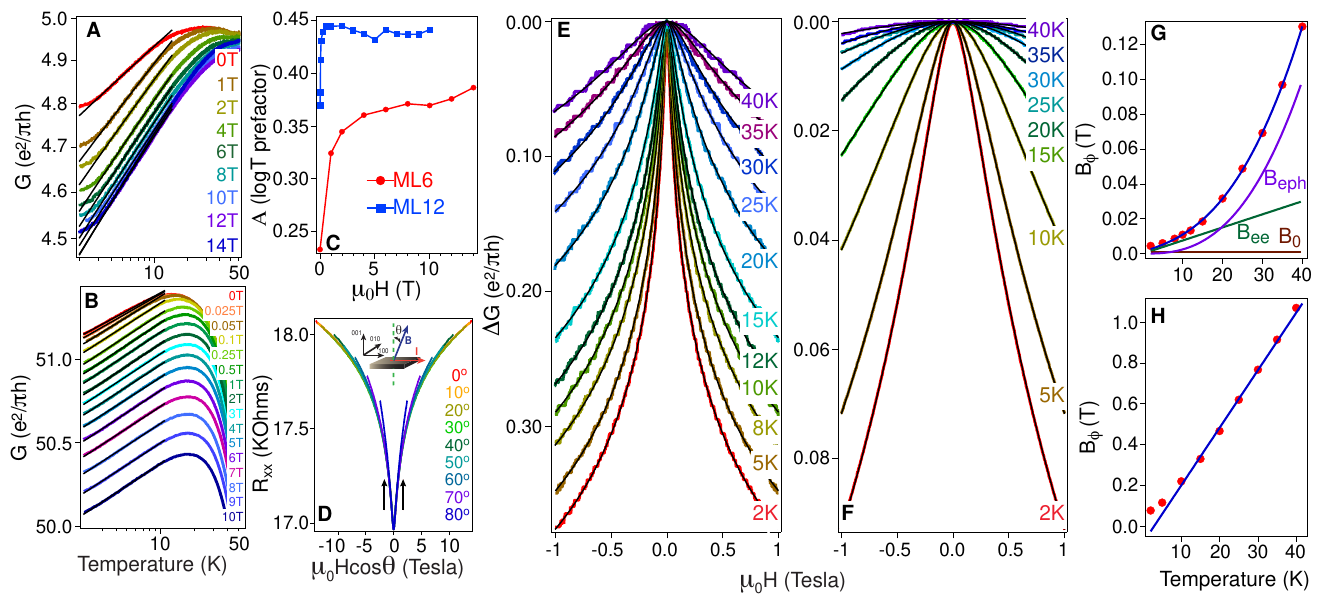}
\caption{\textbf{QI and EEI effects in ultra-thin LuSb/GaSb (001) thin films.} Temperature dependence of conductance in (A) 6 and (B) 12 ML thick films. Logarithmic fits to the conductance are shown in black. (C) Variation of the pre-factor A (see eqn. 2), extracted from the fits in \textbf{A} and \textbf{B}, as a function of applied magnetic field. (D) Resistance of a 6 nm thick film plotted against the perpendicular component of the magnetic field vector for different angle values. The angle is defined in the inset. Arrows indicate that between $\mu_{0}$H = $\pm$ 1 T, magnetoresistance curves obtained at different angular configurations fall on one another establishing the two-dimensional nature of the quantum interference effect. Differential conductance $\Delta$G(B = $\mu_{0}$H) = G(B) - G(0) as a function of magnetic field and the corresponding HLN fits for (E) 12 and (F) 6 ML thick films. Temperature dependence of the extracted dephasing field as a function of temperature for (G) 12 and (H) 6 ML thick films. In \textbf{G}, contributions of B$_{ee}$, B$_{eph}$, and B$_{0}$ to B$_{\phi}$ are shown in green, violet, and brown, respectively.}
\label{fig:QI}
\end{figure*}

\end{document}